\documentclass[final,3p,times,numbers]{elsarticle}
\biboptions{sort&compress}

\usepackage{amssymb}
\usepackage{amsmath}
\usepackage{xr-hyper}
\usepackage{hyperref}
\usepackage{float}
\usepackage{xcolor}
\usepackage{xr} 
\usepackage{caption} 

\usepackage{graphicx}
\usepackage{subcaption}
\journal{arXiv}
\begin{document}
\renewcommand{\figurename}{\textbf{Figure}}
\renewcommand{\thefigure}{\textbf{\arabic{figure}}}

\newcommand{\mpnote}[1]{\textcolor{teal}{[#1]}}

\begin{frontmatter}


\title{ToFiE, a Topology-aware Fiber Extraction workflow for 3D reconstruction of dense and heterogeneous \mbox{biological} fiber networks from microscopy images}

\author[label1,label2]{Risa Togo}
\author[label1]{Sara Cardona}      
\author[label2]{Irène Nagle}
\author[label2]{Gijsje H. Koenderink$^{*}$ }
\author[label1]{Behrooz Fereidoonnezhad$^{*}$ }
\author[label1]{Mathias Peirlinck$^{*}$ }

\affiliation[label1]{organization={Department of BioMechanical Engineering, Faculty of Mechanical Engineering, Delft University of Technology},
                city={Delft},
                country={The Netherlands}}
\affiliation[label2]{organization={Department of Bionanoscience, Kavli Institute of Nanoscience, Delft University of Technology},
            city={Delft},
            country={The Netherlands}}
\cortext[cor1]{Senior authors contributed equally. Correspondence: Gijsje H. Koenderink (G.H.Koenderink@tudelft.nl), Behrooz Fereidoonnezhad (b.fereidoonnezhad@tudelft.nl), Mathias Peirlinck (mplab-me@tudelft.nl).}

\begin{abstract}
Fibrous networks are ubiquitous structural components in biology, spanning cellulose in plant cell walls, fibrin in blood clots, and collagen in the extracellular matrix of animal tissues. Theoretical models predict that network connectivity critically influences their mechanical behavior. However, accurately reconstructing network topology from 3D image data remains a major challenge as current segmentation methods are not designed to preserve network topology and often rely on intensity-based thresholding, which can fragment fibers and distort junction connectivity. Here, we introduce ToFiE, an open-source topology-aware fiber extraction workflow for reconstructing dense and heterogeneous fibrous networks from high resolution microscopy images while preserving connectivity in three dimensions. We validate ToFiE using synthetic fluorescence microscopy images of fiber networks with varying topologies and signal-to-noise ratios. We further demonstrate its performance by reconstructing the fiber networks of a library of collagen gels with various microstructures, imaged using confocal fluorescence microscopy. Altogether, the results establish ToFiE as a practical semi-automated framework for extracting mechanically relevant network information from imaging data across a broad range of fibrous materials. \\

\textbf{Statement of Significance}\\
\\
Fibrous materials in nature derive their function and mechanical properties from their microstructure. Understanding the mechanistic basis of this relation drives research in various fields, including biology, material science, and biomedical engineering, but requires accurate reconstruction of their three-dimensional architecture from images. This remains challenging because segmentation approaches often fail to preserve network topology (connectivity of the fibers) and rely on intensity-based thresholding, making them sensitive to heterogeneity, density, and imaging noise. To address these limitations, we developed a topology-aware fiber extraction workflow and demonstrate robust 3D reconstruction of dense and heterogeneous collagen I networks from confocal fluorescence images. This workflow provides a foundation for more comprehensive studies of the structure, mechanics, and function of a wide range of biological fiber networks. 

\end{abstract}



\begin{graphicalabstract}
\begin{figure}[H]
\centering
\includegraphics[width=\textwidth]{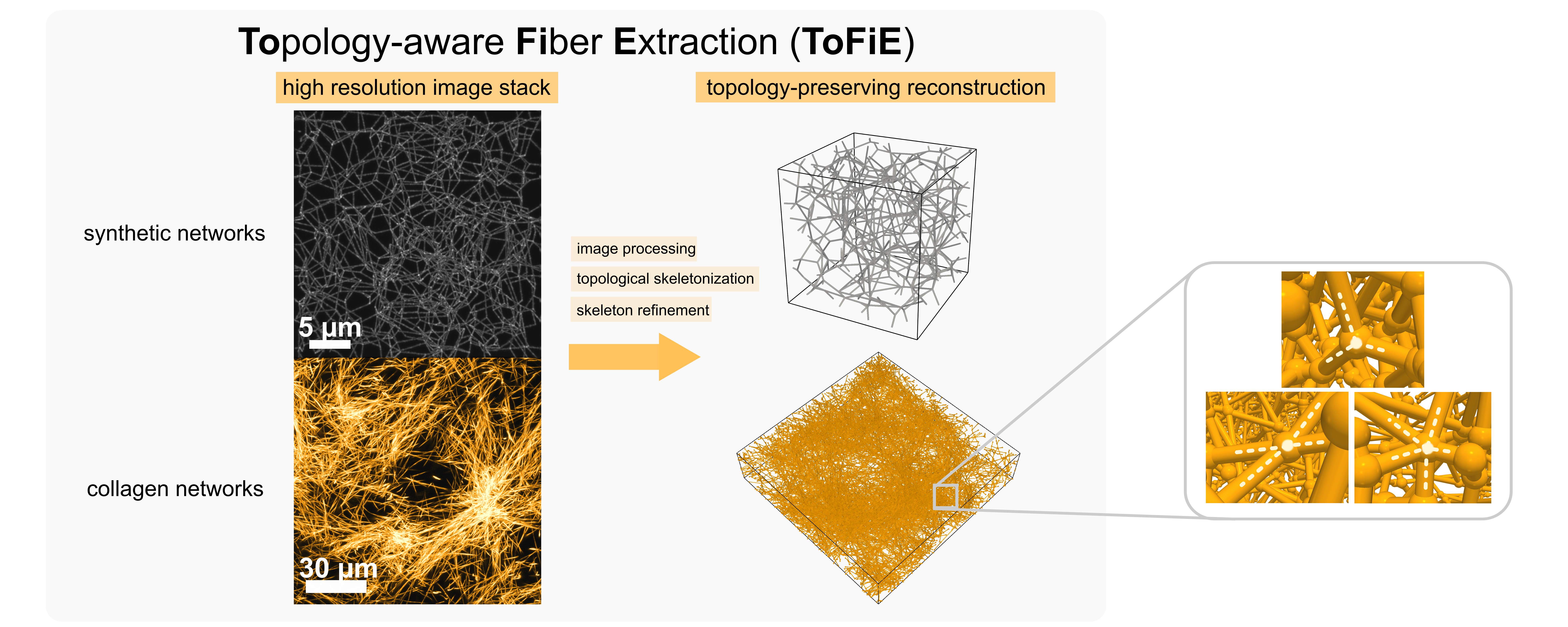}
\end{figure}
\end{graphicalabstract}








\begin{keyword}
3D reconstruction \sep network topology  \sep computational image analysis \sep  confocal fluorescence microscopy \sep tissue microstructure \sep collagen \sep extracellular matrix
\end{keyword}

\end{frontmatter}



\section{Introduction}
\label{Intro}


Fibrous networks are ubiquitous in nature, forming the structural backbone of materials across multiple length scales. In plants, cellulose microfibrils assemble into networks within the cell wall, giving rise to nonlinear mechanical behavior which supports the growth and integrity of plant tissues \cite{Zhang2021MolecularWalls, Chen2025FibrousDevelopment}. In animals, fibrous networks similarly play critical functions. At the cellular level, cytoskeletal networks (actin, microtubules, intermediate filaments) provide structural support, mediate morphological changes and enable migration in response to external or internal cues. The elasticity of actin networks can be controlled by modulating filament branching and capping \cite{Pujol2012ImpactNetworks}, while keratin networks influence mechanotransduction and viscoelasticity of keratinocytes \cite{Ramms2013KeratinsKeratinocytes, Laly2021}. Blood clot mechanics are largely determined by the underlying fibrin network, affecting the success of thrombectomy and thrombolytic treatments \cite{Collet2000InfluenceSpeed,Fereidoonnezhad2021,Domingues2022NanomechanicsFormation,Ramanujam2024RuptureResistance}. Another particularly important example is collagen networks, the main load bearing component of connective tissues. Recent theoretical models predict that the mechanics and fracture of networks of stiff fibers such as collagen are closely linked to their topology, in particular the average network connectivity \cite{Jansen2018TheMechanics., Burla2020ConnectivityFracture,Lindstrom2010BiopolymerProperties, Licup2015StressNetworks,Shivers2019ScalingNetworks, Sharma2016Strain-controlledNetworks}. Across the diversity of fibrous networks, structure is inherently connected to their function. \\
\\
Nevertheless, it remains a significant challenge to quantify the explicit topology of fibrous networks from microscopy images. The challenge can be attributed to the difficulty of resolving network fibers in close proximity as they approach the resolution limit of optical microscopes, and the limitations of current image analysis methods to robustly extract network structures. While the first challenge is inherent to optical imaging methods, the second can be addressed through improved computational approaches. Automated methods to analyse images of fibrous networks fall into two main categories: (i) derivative-based filters or transformations (e.g. Fourier transform, wavelet transform) that extract regionally averaged structural information such as orientation and anisotropy \cite{Alberini2024FourierUnits,vanHaaften2018DecouplingGraft}, and (ii) skeleton-tracing approaches that explicitly reconstruct the network fibers before deriving quantities from it \cite{Xu2015SOAX:Networks, Jaeschke2022Qiber3DanStacks, Bredfeldt2014ComputationalCancer, STEIN2008AnGels, Rossen2021FiberImages}. Tracing methods have an advantage over the former as they can capture the full network description. However, only a few current tools, such as SOAX, CT-FIRE, Qiber3D, and FFA, work in three dimensions \cite{Xu2015SOAX:Networks,Jaeschke2022Qiber3DanStacks, Bredfeldt2014ComputationalCancer,Rossen2021FiberImages,Windoffer2022Quantitative3D}. A key limitation of these and similar tools is the lack of topology awareness embedded within the approach and reliance on intensity-based thresholding, which is highly sensitive to the threshold choice and complicated by signal and structural heterogeneity. Current methods are therefore poorly suited for reconstructing the topology of dense, heterogeneous biopolymer networks displaying a large dynamic range in signal. \\
\\
In this study, we develop ToFiE (topology-aware fiber extraction), a semi-automated workflow that facilitates topology-preserving reconstructions of fibrous networks from image data by utilizing the mathematical definition of topology from Discrete Morse theory (DMT), and relying on a persistence-based thresholding approach to simplify the topology \cite{Sousbie2011TheImplementation}. Together, this approach enables robust reconstruction of junctions even in dense, heterogeneous networks. While this skeletonization strategy was originally developed for studying cosmic filaments \cite{Sousbie2011TheImplementation}, its applicability to biological tissues was recently demonstrated with DISSECT \cite{Merle2023DISSECTEpithelia}. ToFiE extends these approaches in several aspects. ToFiE implements image pre-processing and post-processing refinement steps tailored to noisy biological data, and specifically designed to handle a variety of biological fibrous networks. We first validate the performance of ToFiE on synthetic networks with biologically relevant topology, and then demonstrate its suitability for reconstructing dense and heterogeneous collagen type I networks from 3D confocal fluorescence images. From these reconstructions, we extract several topological and structural parameters including edge length, edge density, orientation, connectivity, and betweenness. Together, these network reconstructions and topological parameters enable direct quantitative analyses of structure–function–mechanics relationships in fibrous materials \cite{Ghodsi2025,Avril2026}.

\section{Materials and methods}
\label{MatMet}

\subsection{ToFiE workflow}

ToFiE is an original workflow with algorithms developed for the pre-processing  of fibrous network images and post-processing of network skeletons. For topology-preserving skeletonization based on DMT and persistent homology, ToFiE interfaces with DisPerSe. DisPerSe is a separate software package by Thierry Sousbie and is licensed independently by its authors \cite{Sousbie2011TheImplementation}. Limited DisPerSe data import helper functions are adapted from the DISSECT framework \cite{Merle2023DISSECTEpithelia}. The workflow consists of three main steps (Figure \ref{figure1}a): (1) image pre-processing to enhance fiber signal and correct imaging artifacts, (2) skeletonization to extract a topology-preserving network representation, and (3) skeleton refinement to obtain biologically meaningful fibers, junction definitions, and network connectivity.

\subsubsection{Image pre-processing}
To address noise in the raw image, we apply a Gaussian filter, followed by a median filter from the scikit-image Python library \cite{vanderWalt2014Scikit-image:Python}. Typical 3D confocal fluorescence images suffer from intensity attenuation with depth due to photobleaching of the dye, light absorption or scattering by the sample, and optical aberrations. The attenuation profile is strongly dependent on the specific sample and imaging conditions and cannot be modeled by a simple decay function. Therefore, we employ a flexible normalization approach as described in Intensify3D \cite{Yayon2018Intensify3D:Stacks}: the contrast and intensities of each z-slice in the image are standardized by re-normalizing pixel intensities falling between a specified lower and upper threshold such that their values span the complete 8-bit range (0–255)  (illustrated in Figure \ref{figure1}b). The normalizing step is important to ensure the algorithm reconstructs in an unbiased manner at all depths. Pixel values outside the thresholds are clipped to the 8-bit range limits. To define the lower limit for intensity normalization, we set it to zero if there is minimal background intensity, otherwise, we manually determine the background intensity for each image stack and apply a constant threshold. The upper threshold value for intensity normalization is determined for each z-slice, using the N-th percentile pixel intensity value as a measure of the overall brightness of the z-slice, where N is a tunable parameter. To ensure continuity between adjacent z-slices, we smooth the N-th percentile intensity value as a function of z using a Gaussian kernel with standard deviation of 10 pixels. The enhanced image stack is deconvoluted with the Gaussian point spread function and the Richardson-Lucy deconvolution algorithm using the SDeconv python framework \cite{Prigent2023SPITFIRe:Videos}, with a padding of 10 pixels at the image boundaries. The resolution of the smoothed image, determined with the Fourier Ring Correlation (FRC) function in the MIPLIB software \cite{Koho2019FourierMicroscopy}, is used as the lateral and axial size of the point spread function (PSF). To enhance the contrast after deconvolution, we re-normalize the image stack to the full 8-bit range.

\subsubsection{Skeleton tracing by discrete Morse theory and persistent homology}
DMT and persistent homology form the mathematical framework for obtaining the initial skeleton of the collagen network from the processed images. For a detailed explanation, we refer the readers to the work of Sousbie \cite{Sousbie2011TheImplementation}. Briefly, DMT partitions an n-dimensional discrete density field into n$-$simplices via Delaunay triangulation and assigns field intensities to the simplices (Figure \ref{figure1}c, top). A three-dimensional image is discretized into points, lines, triangles and triangular pyramids. Then a discrete gradient field is defined within the data, consisting of gradients paths or sequences of simplicial pairs, where each is a gradient pair, and the simplex in the next pair is a facet of the previous simplex. A gradient pair consists of a simplicial pair of $i$-$simplex$ and its lower valued cofacet ($i+1$-$simplex$), or $i$-$simplex$ and its higher-valued facet ($i-1$-$simplex$). Critical simplices, which are not associated with gradient pairs, represent the discrete equivalent of zero-derivative points of the density function in the continuum space. Specifically, a critical $0$-simplex corresponds to a maximum, a critical 1-simplex to a 1-saddle, a critical $2$-simplex to a 2-saddle, and a critical $3$-simplex to a minimum. A topological description of the data is extracted from the gradient vector field linking the critical simplices. The vector paths between the critical 0- and 1- simplex give a skeleton description of the data, also called the discrete 1- manifold. Taking the approach as Merle et al. in DISSECT \cite{Merle2023DISSECTEpithelia}, we trace the fiber skeleton through the discrete 1$-$manifold using the implementation of DMT in the DisPerSe software \cite{Sousbie2011TheImplementation}. The skeleton is defined in filaments, where each filament is described by endpoints and sampling points. However, DMT simply extracts the topological structure of the data, which can include imaging noise. We leverage DisPerSe’s persistent homology algorithm to identify more persistent topological features within the skeleton (Figure \ref{figure1}c, bottom). Persistence is defined as the difference in field intensities of the simplices in a critical pair, a larger difference indicating greater topological importance. By setting a persistence threshold, features associated with lower persistence, such as imaging noise, can be removed. \\
\\
DisPerSe is executed through the Docker image glyg/disperse. In order to reconstruct the synthetic networks and collagen networks, computations were performed on the DelftBlue supercomputer, using 6 CPU cores (Intel Xeon E5-6448Y 2.1 GHz) with 16GB RAM per core \cite{DelftHighPerformanceComputingCentreDHPC2024DelftBlue2}. Four different parameters (cut, smooth, assemble, trimBelow) enabled us to adjust the 1-manifold in DisPerSe (Supplementary Figure \ref{supp1}). The cut parameter sets the persistence threshold. The smooth parameter controls the number of sampling points to average over to smoothen a filament. The assemble parameter defines the maximum angle for merging neighboring filaments in the skeleton. The trimBelow parameter removes topological features associated with an intensity lower than the set threshold. Unwanted connections of fibers to the image background are removed when the threshold is set above the background level. For more details on these parameters, we refer the readers to the DisPerSe manual by Sousbie \cite{Sousbie2011TheImplementation}.

\subsubsection{Skeleton refinement}
\begin{figure}[H]
\centering
\includegraphics[width = \textwidth]{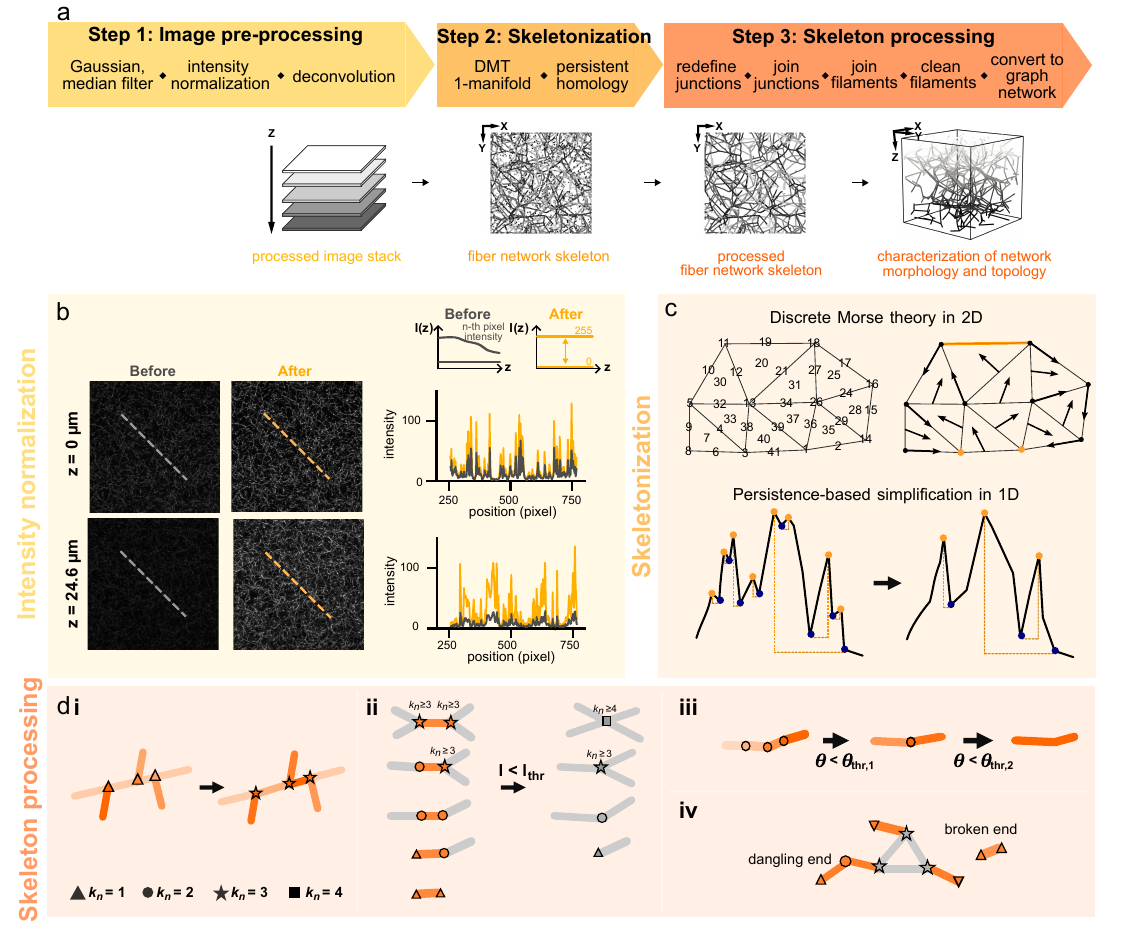}
\caption{\textbf{Overview of the ToFiE workflow.} (a) ToFiE consists of three main steps. Image pre-processing includes denoising, correcting for intensity attenuation with depth, and deconvolution using a theoretical point spread function (PSF). (b) The raw 3D scanning confocal fluorescence image (left) exhibits nonlinear attenuation in intensity with depth. The intensity correction step stretches the pixel intensity values between the upper and lower intensity bound for each z-slice to cover the full 8-bit range. The upper threshold value for intensity normalization is set to the n-th percentile of pixel intensities to capture the overall brightness of each z-slice, while the lower limit for intensity normalization is determined manually based on the background intensity. A fixed lower threshold is applied across the entire image stack. (c) DisPerSe processes the image as a simplicial complex and computes the Discrete Morse-Smale complex, extracting the 1-manifold that connects critical 0- and 1- simplex pairs along discrete gradient paths \cite{Sousbie2011TheImplementation}. Top: Example of 2D data represented by a simplicial complex, adapted from \cite{DGD-DiscretizationinGeometryandDynamics-SFBTransregion109A07Theory}. Each simplex is assigned an intensity value based on the data, and are annotated here to indicate their position in the discrete gradient vector field. Arrows show the discrete gradient flow, and critical 0- and 1- simplices (where the flow ends) are shown in orange. Top: The topology of the 1-manifold is simplified by removing critical 0- and 1-simplex pairs based on their persistence, defined as the intensity difference between the simplexes in a pair. Bottom: Example of persistence-based simplification in one dimension, adapted from \cite{Sousbie2011TheImplementation}. Critical pairs are removed based on their persistence, simplifying the 1-manifold while preserving topologically important structures. (d) Lastly, filaments and junctions of the skeleton are refined in several steps: (i) redefine filaments and junctions to have one filament in between two crosslinks; (ii) remove filaments shorter than a threshold $l<l_{thr}$ to avoid unphysical reconstructions; (iii) merge filaments with an angle smaller than $\theta <\theta_{thr}$; (iv) remove broken ends or dangling ends within the skeleton, and converted into a graph network. 
}
\label{figure1}
\end{figure}

We develop a set of custom functions to further refine filament subunits within the DisPerSe skeleton for the application to biological fibrous networks (illustrated in Figure \ref{figure1}d). First, the original filaments of the skeleton are broken down at the branchpoints or endpoints, such that endpoints cannot be contained within the redefined filaments (Figure \ref{figure1}d-i). This establishes a consistent definition for all filaments within the skeleton. Filaments shorter than a specific length threshold $l_\text{thr}$ are merged with their neighboring filament, or removed, depending on the connectivity $k_n$ of the endpoints of the short filament (Figure \ref{figure1}d-ii). Neighboring filaments that share a similar orientation within an angle threshold $\theta_\text{thr}$ are merged (Figure \ref{figure1}d-iii). To further clean up the skeleton, broken ends are removed. To obtain a fully connected network, dangling ends can also be removed (Figure \ref{figure1}d-iv). The processed skeleton is then converted into an undirected graph with the NetworkX python library \cite{Hagberg2008ExploringNetworkX}, with nodes and edges to represent the endpoints and filaments of the skeleton.

\subsection{Validation of ToFiE with virtual data}
\subsubsection{Simulating confocal fluorescence images of synthetic networks}
\label{Simulating_images}
To validate the ToFiE workflow, we generate synthetic images of biopolymer networks with systematically varying topology and imaging noise, and assess the reconstruction accuracy by comparing to the ground truth. Synthetic networks are created using TopoGEN, explained in detail by Cardona et al. \cite{Cardona2025TopoGEN:Mechanics}. We create two distinct sets of networks: the first set has an average connectivity between three and four, with only branches ($k_n = 3$) and cross-links ($k_n = 4$) as connection types between fibers, reflecting the physiological organization of reconstituted collagen type I networks \cite{Jansen2018TheMechanics.,Burla2020ConnectivityFracture}. The second set of networks has a higher average connectivity between four and five, and connection types ranging from three to ten. This aligns with densely cross-linked fibrin and actin networks, which are associated with higher connectivity \cite{Martinez-Torres2024InterplayNetworks, Spukti2022LargeForces}. Network edge lengths are distributed with a target mean of 2 $\mu$m and a standard deviation of 0.256 $\mu$m \cite{Lindstrom2010BiopolymerProperties,Lindstrom2013Finite-strainNetworks}.\\
\\
From the synthetic networks, we create point cloud representations by sampling each edge at 20 points per $\mu$m. To incorporate heterogeneity in fiber thickness and signal intensity, we additionally sample points along each edge using a uniform random distribution at a resolution of 10 points per $\mu$m, with a 1\% relative uncertainty in its global position. We simulate 3D confocal fluorescence image stacks from the sampled point clouds using the MicroVIP software \cite{Ahmad2021MicroVIP:Platform}. The final cubic image has dimensions of 30 $\mu$m and voxel size of 0.1 $\mu$m. MicroVIP models the confocal microscopy image acquisition process by accounting for various physical factors. We use standard confocal settings for the model similar to the experimental images: an excitation wavelength of 488 nm, an objective magnification of 63x, a numerical aperture of 1.3, a refractive index of the immersion medium of 1.47, and a Gaussian noise standard deviation of 1. To generate images with variable signal-to-noise ratio (SNR) (Section \ref{method_sensitivity_analysis}), we adjust the photon count parameter to the values of 25, 50, 75, 125, 250, 500, and 10000. By setting the Gaussian noise standard deviation and the sampling density of the network, and tuning the photon count parameter, we visually imitate low to high noise levels commonly seen in experimental images (Figure \ref{figure2}a).

\subsection{Sensitivity analysis with synthetic networks}
\label{method_sensitivity_analysis}
We test our workflow performance using the simulated images of fiber networks, generated to match the topology of biological fibrous structures. These serve as the ground truth to assess the accuracy of the 3D reconstructions (see Supplementary Figure \ref{supp1} for ToFiE parameter values used for the reconstruction). A tangible measure of the SNR of the images, defined as the average normalized fiber intensity $\overline I_\text{norm}$ divided by its standard deviation, is calculated with the formula $\text{SNR} = \overline I_\text{norm} /std(I_\text{norm})$ as described in NoiSee \cite{Ferrand2019UsingMicroscopes}. Normalized fiber intensities $I_\text{norm}$ are computed as the difference between the foreground $I_\text{target}$ and background $\overline I_\text{bg}$ intensity values, $I_\text{norm} = I_\text{target} - \overline I_\text{bg}$. We obtain $I_\text{target}$ by applying a mask, generated from the sampled point cloud, to the simulated image. Background intensity values $I_\text{bg}$ are obtained by applying the inverse of the morphological dilated mask to the image, which excludes fiber regions and ensures that only background signal is measured. We study two structural parameters, edge lengths $l$, and azimuthal angles $\theta$, defined as the in-plane edge angle measured counter-clockwise from the positive x-axis. To evaluate the accuracy of our reconstructions, two distinct metrics are leveraged: the Kullback$-$Leibler divergence (KLD) score and the recall score. KLD quantifies the difference in information represented by the probability distribution of the reconstructed and ground truth (GT) network. The probability distribution of edge lengths $l$ is created as a normalized histogram with 30 equally spaced bins for $l \in [0,15]$ $\mu\text{m}$. Similarly, the probability distribution of $\theta$ is created as a normalized, edge-length-weighted histogram, with 18 equally spaced bins for $\theta\in [-\frac{\pi}{2},\frac{\pi}{2}]$. Then we compute the corresponding KLD according to $\text{KLD}=\sum_i P_r \log( \frac{P_r}{P_{GT}})$, where $i$ represents an event (e.g., obtaining an edge length between 1 $\mu$m and 1.5 $\mu$m), and $P_r$ and $P_{GT}$ are the corresponding probabilities of the reconstruction and GT network. With the recall score, we locally quantify the spatial overlap between the reconstructed and GT nodes by computing confusion matrix components. Due to the discretization of the networks for confocal image simulation, reconstructed networks are not perfectly aligned with the GT. Therefore, we consider two nodes to overlap if the distance between their centers is less than $2r$, with $r=$ 0.5 $\mu$m. We select this value because the recall score increases with $r$ and stabilizes first at around 0.5 µm (Supplementary Figure \ref{supp2}). To avoid reconstruction artifacts, we exclude nodes less than 1 $\mu$m away from the domain boundaries, which corresponds to roughly half the average fiber length (Supplementary Table \ref{stable1}), ensuring that partially imaged fibers and boundary conditions of DMT do not bias the analysis. Nodes that overlap and have the same $k_n$ are counted as true positives (TP). GT nodes with no overlapping reconstruction with the same $k_n$ are false negatives (FN), and reconstructed nodes with no overlapping GT node of the same $k_n$ are false positives (FP). We define the recall score as the total number of correctly reconstructed nodes divided by the total number of GT nodes: $recall =\frac{TP}{TP+FN}$.

\subsection{Application of ToFiE to experimental data}
\subsubsection{Collagen gel preparation}
Collagen gels are prepared at different polymerization temperatures to produce networks with distinct topologies \cite{Jansen2018TheMechanics.}. DyLight 550 NHS Ester (ThermoScientific)-labeled collagen solution is prepared from bovine atelocollagen type I solution (5133, Biomatrix Fibricol, Advanced BioMatrix) using a protocol adapted from the work of Remy et al. \cite{Remy2023InvadopodiaMatrices} and Doyle et al. \cite{Doyle2018FluorescentImaging.} (for details see Supplementary Information). The samples for confocal fluorescence imaging are prepared on ice by neutralizing bovine atelocollagen type I solution with 1:2 mass fraction of fluorescent-labeled and unlabeled collagen monomers, in one part 10x PBS (pH 7.4, Gibco), 0.1 M NaOH, and MilliQ water to reach the target collagen concentration (1.5, 2.5, or 3.5 mg/mL) and pH 7.4. The samples are pipetted into an 18-well glass bottom $\mu$-slide (Ibidi, 81817), which is kept in a temperature regulated incubator (Thermomixer Comfort, Eppendorf) for at least 90 minutes at 37$^{\circ}\text{C}$, or overnight at 26$^{\circ}\text{C}$, until complete polymerization of the networks. To enhance heat transfer and maintain a humid environment, a drop of water is added between the bottom of the slide and the heated plate, and neighbouring wells in the slide are filled with deionized water. Before imaging, an oxygen scavenger mix consisting of 1 mM Protocatechuic acid (PCA, PHL89766) and 0.05 $\mu$M Protocatechuate-3,4-dioxygenase (PCD, ICNA0215197505) \cite{Aitken2008AnExperiments.} in 1x PBS is pipetted on top of the polymerized gels.

\subsubsection{Confocal fluorescence imaging of collagen gels}
Fluorescence image stacks are acquired using a Stellaris FALCON 8 confocal microscope (Leica Microsystems) equipped with a 63x glycerol-immersion objective (HC PLAN APO CS2, NA 1.3, Leica Microsystems) and a white laser ($80$ MHz) controlled by an acousto-optical beam splitter. Samples are excited at $553$ nm with a laser intensity of 5\%, a line scan speed of $40$-$120$ Hz, and two line accumulations. Emission light is collected between $561$ nm and $740$ nm using a HyD X detector set to a gain of $20$. The reference plane ($z=0$) is defined at the glass slide, whose position is determined by the bright reflection signal. Image stacks are acquired starting $20$ $\mu$m above the slide (corresponding to approximately ten average fiber lengths, Supplementary Table \ref{stable1}) to avoid boundary effects on the imaged network structure, with an isotropic voxel size of $110$ nm and dimensions of $112$ x $112$ x $25$ $\mu m^3$ (X x Y x Z).

\subsubsection{Analysis of reconstructed collagen networks}
The confocal images of the collagen networks are processed and reconstructed using ToFiE (Supplementary Figure \ref{supp1} for ToFiE parameter values used for the reconstruction). We quantify three additional network-related parameters: the edge density, defined as the total edge length per unit volume; the network heterogeneity, defined as the standard deviation in edge density across $n$ non$-$overlapping equal partitions of the reconstruction; the normalized betweenness centrality of node $u$ in the network, defined as the ratio of the number of shortest paths between all pairs of nodes passing through $u$ (weighted by edge lengths) and the number of shortest paths between all possible node pairs (weighted by edge lengths). To validate whether the reconstructions preserve network topology, 50 randomly selected junctions per reconstruction are manually inspected by two independent observers, and compared to those observed in the image. The junction accuracy score (JAS) is then computed as the fraction of correctly reconstructed junctions.

\newpage

\section{Results}

\subsection{Topology-preserving synthetic network reconstructions}
\begin{figure}[H]
\centering
\includegraphics[width = \textwidth]{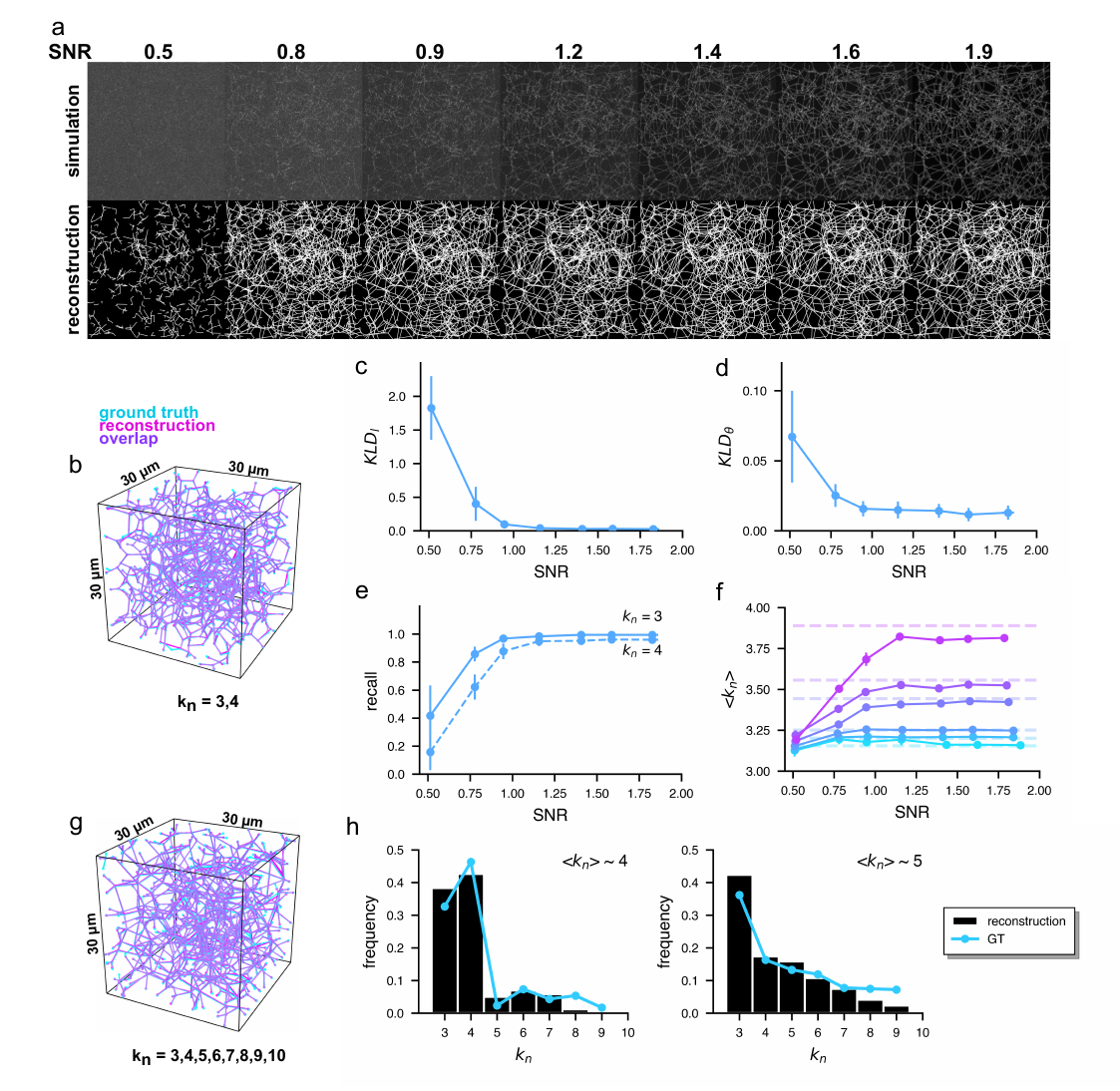}
\caption{\textbf{Topology-preserving reconstruction of simulated fibrous networks.} (a, top) Maximum intensity z-projections of simulated confocal fluorescence images of a synthetic network at varying signal-to-noise ratios (SNR). (a, bottom) Corresponding z-projections of the reconstructed networks. (b,g) 3D visualization of the reconstruction (magenta) obtained at the highest SNR condition tested ($\approx 1.9$), superimposed with the ground truth (GT) synthetic network (cyan). (c,d) Kullback-Leibler divergence (KLD) quantifying the similarity between the GT network and reconstructions for (c) edge length $l$ distributions, and (d) azimuthal angle $\theta$ distributions as a function of image SNR. Error bars represent the standard deviation across $N=15$ independent networks. (e) Recall score for nodes with connectivity $k_n=3$ (solid line) or $k_n=4$ (dashed line). Recall is defined as recall $= \frac{TP}{TP+FN}$, quantifying the overlap of nodes with the same connectivity between the reconstructed and GT networks. Nodes that overlap and have the same $k_n$ are counted as true positives (TP), and GT nodes with no overlapping reconstructed node of the same $k_n$ are false negatives (FN). Error bars represent the standard deviation across $N=15$ independent networks. (f) Average connectivity $<k_n>$ of reconstructions (solid line) as a function of image SNR, for networks with true average connectivity (dashed line) ranging from $3$ to $4$. Error bars represent the standard deviation across $N=2$-$3$ independent networks. (h) Comparison of the connectivity distribution between the GT network and corresponding reconstruction (obtained at the highest SNR condition using the length threshold of $0.85$ µm) for a network with $<k_n>\approx 4$ (left), or $<k_n> \approx 5$ (right).}
\label{figure2}
\end{figure}

To address the lack of topology-awareness in image segmentation approaches, we developed ToFiE, a semi-automated workflow for the 3D reconstruction of fibrous networks from confocal image stacks. We first test ToFiE on synthetic networks that recapitulate the topology of \textit{in vitro} reconstituted collagen I networks \cite{Jansen2018TheMechanics.,Burla2020ConnectivityFracture}. These networks have an average connectivity, $<k_n>$, between three and four and have a combination of junctions with branches ($k_n = 3$) and cross-links ($k_n = 4$). We create artificial confocal fluorescence images of the synthetic networks with different SNR to reflect experimental conditions (Figure \ref{figure2}a, top). In the highest SNR condition, Poisson and Gaussian noise are minimized, and signal heterogeneity is present due to the sampling strategy, which introduces uncertainty by randomly sampling points along each fiber (Section \ref{Simulating_images}). Applying ToFiE to these artificial images yields 3D reconstructions that closely overlap with the GT network (Figure \ref{figure2}b,g). For a global comparison, we quantify the similarity between reconstructions and GT networks using the KLD of the edge length $l$ and azimuthal angle $\theta$ distributions (Figure \ref{figure2}c,d, Supplementary Figure \ref{supp4}). Lower KLD scores indicate higher similarity between the reconstruction and GT. For SNR$>$ 0.75, the KLD score of the edge length distributions remains low (about 0.1). For lower SNR the KLD score increases to about 1.8 (Supplementary Figure \ref{supp4}), as the reconstruction accuracy drops because of fragments being traced out instead of full length fibers. Compared to the edge lengths, $\theta$ distributions are less sensitive to the images' SNR (Figure \ref{figure2}d), with KLD score increasing slightly to $0.07$ at the lowest SNR condition. Overall, this supports the robustness of identified fiber orientations and fiber lengths at higher SNR. \\
\\
To evaluate whether ToFiE reproduces the topology of the synthetic networks, we assess the one-to-one correspondence of nodes between the GT network and reconstruction, by computing the recall score (the proportion of node junctions accurately captured by the reconstruction, ranging from 0 to 1). A perfect reconstruction corresponds to a recall score of one, indicating complete overlap between reconstructed and GT nodes. Recall scores are computed for $k_n$=3 and $k_n$=4 nodes and allowing a small positional tolerance in the node overlap to account for slight spatial misalignment of the GT network and reconstruction. The recall of $k_n$=3 and $k_n$=4 nodes is high, reaching 0.99 and 0.95, respectively, and decreases only when SNR goes below 1 (Figure \ref{figure2}e). Thus, the reconstructions are able to capture 3-junctions slightly more reliably than 4-junctions. Consistent with this, as the true connectivity increases (i.e., the network contains more 4-junctions), the systematic underestimation of $<k_n>$ also increases (Figure \ref{figure2}f). Nevertheless, within the range of $<k_n>$ $\in [3.15,3.9]$, differences in $<k_n>$ remain distinguishable between the networks.\\
\\
Biopolymer networks such as fibrin and actin can harbour junctions with higher connectivity ($k_n>4$), depending on the crosslink density \cite{Martinez-Torres2024InterplayNetworks} and molecular crowding \cite{Spukti2022LargeForces}. Therefore, we extend the analysis to synthetic networks with higher average connectivity ($<k_n> \approx 4-5$), and junctions ranging from $k_n$= 3 to 10 (Figure \ref{figure2}g). ToFiE is also able to capture higher connectivity junctions, as shown with the closely correlated connectivity distributions of the reconstruction and GT network (Figure \ref{figure2}h, Supplementary Figure \ref{supp5}) and high recall score for $k_n=3-6$ (Supplementary Figure \ref{supp6}). However, the recall drops notably for $k_n>6$. We attribute this weakness mainly to the sampling strategy and image quality of the simulations, but it also suggests a decreased performance for reconstructing higher connectivity networks.

\subsection{Topology-aware collagen I network reconstructions}

To demonstrate the application of ToFiE to experimental data, we reconstruct collagen networks from fluorescence images. Collagen gels with varying microstructure are prepared by adjusting the concentration (1.5, 2.5, or 3.5 mg/mL) and polymerization temperature (26$^{\circ}\text{C}$ or 37$^{\circ}\text{C}$). The gels are imaged using a scanning confocal microscope, with a voxel size of 0.11 µm across a z-depth of 24.6 µm to capture the three-dimensional network structure. At 37$^{\circ}\text{C}$, collagen forms a homogeneous mesh whose density increases with concentration (Figure \ref{figure3}a). The networks polymerized at 26$^{\circ}\text{C}$ display concentration-dependent bundling: at the concentration of 1.5 mg/mL, the networks are largely homogeneous, while at the concentration of 3.5 mg/mL, the networks consist of dense regions of fiber bundles and sparse regions with few fibers. Fibers also appear thicker at lower temperatures, translating to higher fluorescence signals.\\
\\
To explore whether these reconstructions capture structural changes across conditions, we quantify the edge density, defined as the total edge length per unit volume, where edge length is the inter-crosslink fiber distance (Figure \ref{figure3}b, Supplementary Figure \ref{supp7}). Consistent with the qualitative observations, we find that at 37$^{\circ}\text{C}$, the edge density increases with collagen concentration (0.33 µm$^{-2}$ and 0.43 µm$^{-2}$ at 1.5 and 3.5 mg/mL respectively), whereas at 26$^{\circ}\text{C}$ the average edge density slightly decreases (0.38 µm$^{-2}$ and 0.28 µm$^{-2}$ at 1.5 and 3.5 mg/mL, respectively). Visual inspection of the images further suggests increased heterogeneity at lower polymerization temperatures and higher collagen concentrations (Figure \ref{figure3}a). To quantify spatial heterogeneity, each reconstruction is subdivided into $n$ non-overlapping equal partitions in the xy-plane. The standard deviation in edge density across the partitions is computed as a measure of the heterogeneity (Figure \ref{figure3}c). Networks polymerized at 37$^{\circ}\text{C}$ exhibit a high degree of homogeneity over the length scales analyzed, with consistently low standard deviations across partition sizes and collagen concentrations. In contrast, the networks polymerized at 26$^{\circ}\text{C}$ show a pronounced increase in standard deviation starting at just 4 partitions (corresponding to a length scale of 56 µm). The 3.5 mg/mL network polymerized at 26$^{\circ}\text{C}$ is especially heterogeneous, with standard deviations approaching 0.2 µm$^{-2}$.\\
\begin{figure}[H]
\centering
\includegraphics[width = \textwidth]{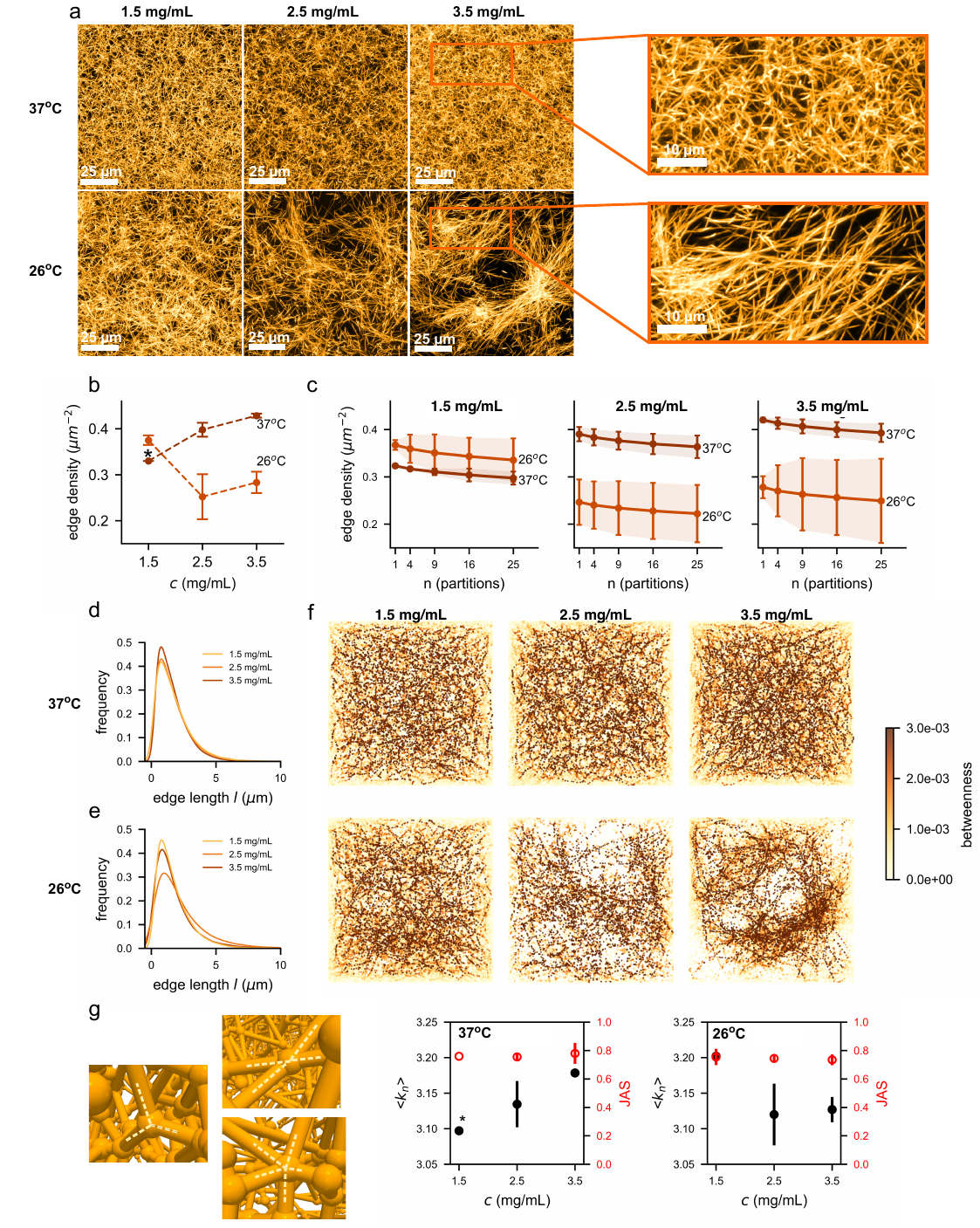}
\end{figure}
\captionof{figure}{\textbf{Topology-aware reconstruction of dense and heterogeneous collagen I networks.} (a) Maximum intensity z-projections of preprocessed confocal fluorescence images of collagen I gels (concentrations: 1.5, 2.5, 3.5 mg/mL, polymerization temperatures: 37$^{\circ}\text{C}$ and 26$^{\circ}\text{C}$), projected over a depth of 4.95 $\mu$m. (b) Average edge density as a function of the collagen concentration, at 37$^{\circ}\text{C}$ and 26$^{\circ}\text{C}$. (c) Average edge density as a function of the number of equally sized partitions $n$ of the reconstruction. Error bars represent the standard deviation between partitions, averaged over N=2 independent networks (* show N=1 network). (d,e) Pooled kernel density estimates of edge lengths for collagen concentrations of 1.5, 2.5, 3.5 mg/mL, at (d) $37^oC$ and (e) $26^oC$. (f) Plot of network nodes colored by their betweenness centrality values. (g) Average connectivity $<k_n>$ of the reconstructed networks. JAS is computed from manual inspection of 50 random junctions per reconstruction by two independent observers, as the fraction of correctly reconstructed junctions. Error bars represent the standard deviation between N=2 independent networks.\\ }
\label{figure3}

\begin{figure}[H]
\centering
\includegraphics[width = \textwidth]{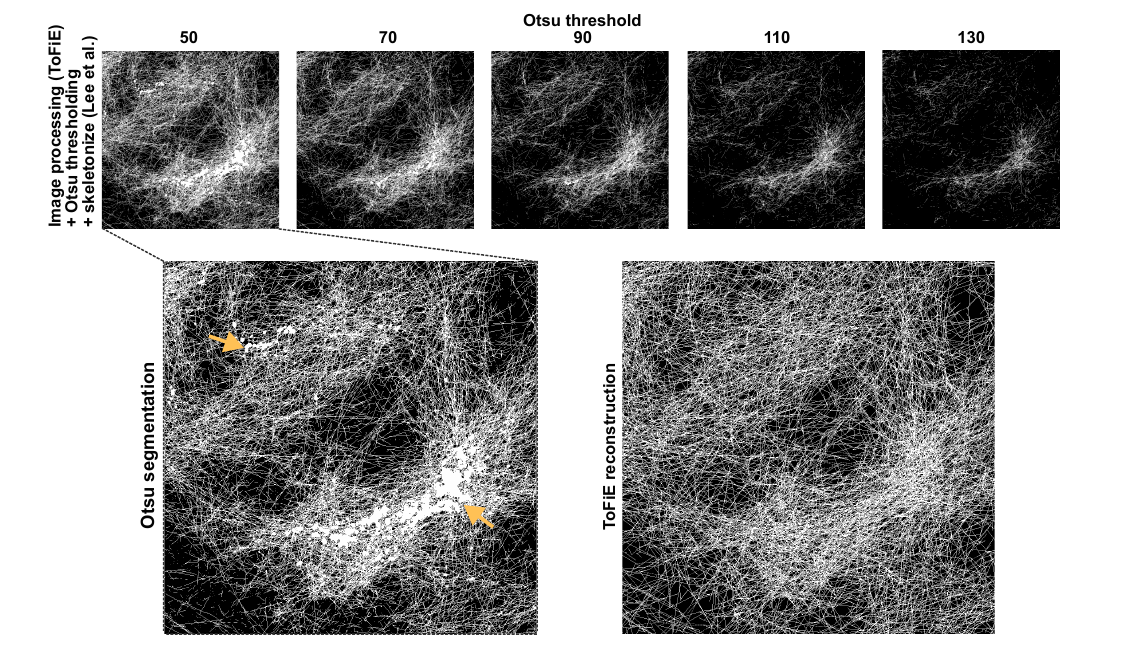}
\caption{\textbf{Comparison of Otsu segmentation and ToFiE reconstruction of a dense and heterogeneous collagen network (3.5 mg/mL, polymerized at 26$^{\circ}\text{C}$).} Maximum intensity z-projections are shown for the Otsu-thresholding based skeleton (top row, bottom left) and the ToFiE reconstruction (bottom right). For Otsu segmentation, Otsu thresholding is applied in ImageJ to the preprocessed confocal image (after initial ToFiE image processing step), using a fixed upper threshold value of 255 and varying lower thresholds (50, 70, 90, 110, or 130). The resulting binary mask is thinned to a skeleton using Skeletonize3D in ImageJ \cite{Lee1994BuildingAlgorithms}. Increasing the lower threshold removes low-intensity pixels indiscriminately, regardless of whether the pixel is part of the background or in the middle of a fiber. This leads to fragmented fibers and a loss of the global network topology. Decreasing the lower threshold results in poorly defined junction segmentation, leading to artifacts in the skeleton that appear as bright spots in the z-projection, as indicated by the orange arrows.}
\label{figure4}
\end{figure}

We further analyze network topology using edge lengths and connectivity. The distribution of edge lengths (pooled kernel density) across two samples for networks polymerized at 37$^{\circ}\text{C}$ and 26$^{\circ}\text{C}$ are shown in Figure \ref{figure3}d,e. At 37$^{\circ}\text{C}$, the mean edge length decreases consistently from 2.13 µm to 1.83 µm with increasing concentration (Figure \ref{figure3}d, Supplementary Figure \ref{stable1}). In contrast, at 26$^{\circ}\text{C}$ the trend is non-monotonic: the mean edge length slightly increases from 2.01 µm to 2.39 µm when raising the concentration from 1.5 mg/mL to 2.5 mg/mL, while it decreases to 2.12 µm at 3.5 mg/mL. The edge length distributions of the 1.5 mg/mL and 3.5 mg/mL network polymerized at 26$^{\circ}\text{C}$ are not significantly different (p-value $>$0.05). Another structural parameter that can be explored using the reconstructions is node centrality, quantified by node betweenness centrality, a graph measure of how nodes are physically linked. Nodes with high betweenness act as influential bridges whose disruption can impact stress transmission. High betweenness nodes are enriched along collagen fiber bundles, while being more randomly distributed in the homogeneous networks (Figure \ref{figure3}f). Globally, the average connectivity of the networks polymerized at 37$^{\circ}\text{C}$ increases monotonically with concentration (3.09 to 3.18, Figure \ref{figure3}g), while networks polymerized at 26$^{\circ}\text{C}$ show a non-monotonic decrease (3.20 to 3.13 for 1.5 mg/mL and 3.5 mg/mL collagen networks respectively), reflecting the bundling-induced structural heterogeneity. To validate these reconstructions, 50 random junctions within each network are manually assessed, obtaining a junction accuracy score of $\approx 0.75$ across conditions. These results support that the ToFiE reconstructions preserve the topology of dense and heterogeneous networks, and capture features relevant for mechanical studies.\\
\\
Finally, we compare our ToFiE reconstruction to a standard intensity-based image segmentation approach, Otsu segmentation. Otsu segmentation is carried out by binarizing the preprocessed image using Otsu thresholding, followed by skeletonization of the resulting mask using Lee's thinning algorithm in ImageJ \cite{Lee1994BuildingAlgorithms}. Reconstructing the same heterogeneous collagen network (3.5 mg/mL polymerized at 26$^{\circ}\text{C}$) with Otsu segmentation proves difficult in densely bundled regions (Figure \ref{figure4} , Supplementary Figure \ref{supp8}, Videos \ref{suppvideo1}, \ref{suppvideo2}, and \ref{suppvideo3} ). Higher Otsu thresholds resolve bright fiber bundles but lose dim fiber pixels, retrieving a fragmented skeleton. Lowering the Otsu threshold allows segmentation of the whole network to a similar extent as ToFiE, however, bright artifacts emerge in bundled regions indicating that topology cannot be uniquely defined from the binary mask. In contrast, ToFiE reconstructs a connected network across both dense and sparse regions without bright artifacts and remains robust for different persistence thresholds (Figure \ref{figure4}, Supplementary Figure \ref{supp9}).

\section{Discussion}
In this work, we develop, validate, and demonstrate the semi-automated ToFiE workflow for topology-aware 3D reconstruction of fibrous networks from high resolution microscopy images. The first step of our workflow addresses noise, intensity attenuation with depth, and optical distortions in the raw confocal fluorescence images. By enhancing the image signal in this step, we improve the performance of discrete Morse theory and persistent homology to trace the fiber network skeleton. The topology of the 1-manifold is defined mathematically, and is accurate to the underlying structure given an optimal image. Using this mathematical definition, DisPerSe has been applied by Merle et al. in DISSECT for segmenting apical cell surfaces and their cytoskeleton \cite{Merle2023DISSECTEpithelia}. In this study, we identify image processing (Figure \ref{figure1}b), the breakdown of filaments at junctions, and the redefinition of individual filaments within the DisPerSe skeleton (Figure \ref{figure1}c) as crucial steps for accurately retrieving the topology of biological networks, and integrate it into the ToFiE workflow. The workflow is semi-automated, meaning that each step requires optimization by the user for their particular dataset (Supplementary Figure \ref{supp1}).\\
\\
To evaluate the performance of ToFiE, we generated a set of artificial confocal images of synthetic networks with known underlying topology, which serve as ground truth for validation (Figure \ref{figure2}). Characteristics of experimental images are simulated by sampling synthetic networks using a uniform distribution with positional uncertainty to create signal heterogeneity along fibers, and by varying the photon count parameter in the MicroVIP software to change the SNR \cite{Ahmad2021MicroVIP:Platform}. We show that network morphological parameters such as edge lengths and azimuthal edge angles are accurately reconstructed from these images over a wide range of SNR. We also demonstrate that topology is preserved in the reconstructions with high accuracy (quantified by the recall score). However, the recall of junctions with higher nodal connectivity ($k_n >6$) diminishes as shown with the second set of synthetic networks (Supplementary Figures \ref{supp5}, \ref{supp6}). We attribute this to three aspects of our approach: (i) the quality of the simulated images, which exhibit abrupt intensity changes from pixel to pixel, (ii) the sampling strategy, which samples junctions as many times as the number of edges connected to it – so a more connected junction becomes oversampled compared to edges of the network, and (iii) the discretization of the sampled point cloud into finite pixels to compute the image. We show that by carefully calibrating the workflow, in particular increasing $l_{thr}$, we can improve the recall of junctions $k_n >6$ (Supplementary Figure \ref{supp6}), through merging of junctions that are close to each other. This also implies that topological information is retained in the reconstructions. For the application of ToFiE to real images, better performance is expected as the limitations described above are less relevant. Our findings nevertheless highlight the importance of selecting a suitable $l_{thr}$ for the particular fibrous network of interest.  \\
\\
Previous works showed that collagen concentration and polymerization temperature alter the architecture of collagen networks by changing the nucleation and growth rate of fibers \cite{Jansen2018TheMechanics.,Cooper1970, Jones2014TheMatrix, Morozova2018ElectrostaticFormation}. We exploit this sensitivity here to create dense and heterogeneous collagen networks. Such networks are challenging structures to reconstruct with conventional image segmentation approaches due to the large dynamic range in fiber intensity and tightly packed fibers within bundles. However, accurate reconstruction is essential for understanding connective tissue mechanics \cite{Domingues2022NanomechanicsFormation,Parry1988TheTissue,Deeken2017MechanicalRepair}, and mechanical interactions between cells or multicellular systems with collagen \cite{Nagle2025InvasiveAdhesion, Kim2017Stress-inducedNetworks, Li2021BiophysicalBehaviors, Han2016OrientedIntravasation, Han2018CellMatrix., Mark2020CollectiveNetworks}. Therefore, using our topologically varying, rich experimental dataset, we assess ToFiE performance across different collagen microstructures. At the polymerization temperature of 37$^{\circ}\text{C}$, ToFiE reconstructions measure a consistent decrease in edge length and an increase in edge density with increasing collagen concentration (Figure \ref{figure3}b,d, Supplementary Table \ref{stable1}). At 26$^{\circ}\text{C}$, edge lengths show no clear trend with concentration, and the edge density decreases with concentration (Figure \ref{figure3}b,e), which may be attributed to increased fibril bundling. The mean edge length ranges from 1.8 – 2.4 µm, which is consistent with the work of Lindstrom et al., who found $\overline{l}$ values ranging from 1.3 – 2.0 µm for bovine collagen I at the concentration of $1–4$ mg/mL \cite{Lindstrom2013Finite-strainNetworks}. Edge lengths closely follow a log-normal distribution (Supplementary Table \ref{stable1}) suggesting that network fibers are reliably segmented as similar log-normal length distributions have been found in earlier works \cite{Lindstrom2010BiopolymerProperties,Lindstrom2013Finite-strainNetworks}. \\
\\
We compared our ToFiE collagen network reconstruction with an intensity-based image segmentation approach (Otsu thresholding followed by Lee's skeleton thinning algorithm) \cite{Lee1994BuildingAlgorithms}. As expected, the conventional Otsu segmentation approach struggles to simultaneously reconstruct sparse fiber regions without introducing artifacts in the brighter bundled fiber regions (Figure \ref{figure4}). State-of-the-art approaches such as Qiber3D, and FIRE, which work by morphological thinning or tracing of the Euclidean distance map after image binarization \cite{Jaeschke2022Qiber3DanStacks,Bredfeldt2014ComputationalCancer,STEIN2008AnGels}; and FFA, which traces fibers stepwise based on a user-defined intensity threshold \cite{Rossen2021FiberImages}, face the same limitations because of the reliance on absolute intensity thresholding. While ToFiE similarly depends on the persistence threshold to balance under- or over tracing the network, the persistence threshold and Otsu threshold are qualitatively very different in their workings (Supplementary Figure \ref{supp9}): a lower persistence threshold (over tracing) still fully resolves fiber bundles without artifacts, and a higher persistence threshold (under tracing) retrieves a simplified but connected network topology. \\
\\
Theoretical models predict that the mechanics of stiff fiber networks like collagen are closely linked to their average network connectivity \cite{Jansen2018TheMechanics.,Lindstrom2010BiopolymerProperties, Licup2015StressNetworks,Shivers2019ScalingNetworks, Sharma2016Strain-controlledNetworks}. Verifying these predictions experimentally has been challenging and has so far relied on indirect methods involving fitting of experimental rheology data to fiber network simulations to obtain the average network connectivity \cite{Jansen2018TheMechanics.,Burla2020ConnectivityFracture}. Using ToFiE, it is possible to directly test the relation between connectivity and stiffness or fracture strength. Additionally, simulations frequently study the behavior of 2D or 3D lattice-based networks with varying $<z>$ that are entirely homogeneous \cite{Licup2015StressNetworks,Shivers2019ScalingNetworks,Lerner2024EffectsNetworks}. Little is known about how heterogeneity in disordered networks influences their mechanical behavior \cite{Hatami-Marbini2009EffectNetworks}. From visual inspection, heterogeneity introduces additional length scales to the collagen network, within the bundles and between bundles (Figure \ref{figure3}a,c), which are not captured by edge length measurements (Figure \ref{figure3}d,e, Supplementary Table \ref{stable1}). ToFiE reconstructions enable quantifying local differences in the connectivity and length scales associated with bundling, and may provide insights into how a network differentially responds to microscopic loads (such as forces applied by cells) versus macroscopic loads.\\
\\
ToFiE is not limited to tracing collagen networks or to confocal fluorescence images. Many other biopolymers in nature form similar disordered fibrous networks with mechanically important roles: fibrin networks are the load-bearing and stabilizing components of blood clots \cite{Janmey1983RheologyDeformations, Kim2014StructuralCompression}; actin, intermediate filaments and microtubules form composite cytoskeletal networks that jointly regulate cellular shape and mechanics \cite{Conboy2024HowStudies}. The workflow can reconstruct any dense biological structure given a high resolution 2D or 3D image. If topology (i.e. connectivity) is not essential to quantify, ToFiE can also be applied to label-free imaging modalities such as confocal reflectance and second-harmonic generation images, popular with collagen \cite{Nagle2025InvasiveAdhesion, Yang2009RheologySelf-Assembly}, to extract morphological parameters (in-plane angle, edge length, and edge density) with high accuracy. ToFiE can even be applied to atomic force microscopy images of nanoscale fibrous structures, as the 3D topographical data share similar signal and contrast profiles with confocal microscopy images. The mechanical relevance of ToFiE reconstructions makes integration with traction force microscopy another promising direction for investigating how cell-generated traction forces are transmitted through heterogeneous and discrete fibrous networks in the local cell microenvironment \cite{Barrasa-Fano2021TFMLAB:Microscopy, BarrasaFano2025}. The current workflow does not incorporate a tracking algorithm to link reconstructions across time steps, however this would be a useful extension that can enhance analyses of network structures under macroscopic deformation or cell-mediated remodeling \cite{Kang2009NonlinearGels, Arevalo2015StressMicroscopy, Kim2023MechanicalExperiments, Bohringer2023FiberContractility}.

\section{Conclusions}
ToFiE is a topology-aware image segmentation tool designed to extract the three-dimensional structure of biopolymer networks from microscopy images. We show that ToFiE accurately reconstructs both the morphology (in-plane angle, edge length, and edge density) and topology of synthetic networks resembling fiber networks formed by collagen, actin, and fibrin. Our reconstructions of collagen type I networks from 3D confocal fluorescence images represent a significant advancement in experimental efforts to extract collagen network connectivity directly from images and enable us to quantify local and global structural changes in the networks (edge
length, edge density, orientation, connectivity, and betweenness) across varying collagen concentrations and polymerization temperatures. We envision that ToFiE will facilitate quantitative analyses of a wide range of biopolymer network topologies and offer deeper insights into structure-mechanics relationships.

\section{Credit authorship contribution statement} 
\textbf{Risa Togo:} Writing – Original Draft, Visualization, Validation, Software, Methodology, Investigation, Formal analysis, Data curation, Conceptualization. \textbf{Sara Cardona:} Writing – review \& editing, Supervision, Validation, Methodology, Conceptualization. \textbf{Ir}\textbf{\`{e}}\textbf{ne Nagle:} Writing – review \& editing, Supervision, Validation, Resources, Methodology, Conceptualization.
\textbf{Gijsje Koenderink:} Writing – review \& editing, Validation, Supervision, Project administration, Funding acquisition, Conceptualization.
\textbf{Behrooz Fereidoonnezhad:} Writing – review \& editing, Validation, Supervision, Project administration, Funding acquisition, Conceptualization.
\textbf{Mathias Peirlinck:} Writing – review \& editing, Supervision, Project administration, Funding acquisition, Conceptualization.

\section{Declaration of competing interest}
The authors declare that they have no known competing financial interests or personal relationships that could have appeared to influence the work reported in this paper.

\section{Acknowledgements}
We thank Imke Jansen, Emma Hazekamp, and Frank Gijsen (Erasmus MC) for sharing imaging data that were used in the initial phase of developing the ToFiE workflow. We thank Sebastien Callens (Eindhoven University) for useful discussions regarding image processing. We thank Jorge Barrasa-Fano (KU Leuven) for his valuable assistance in testing the ToFiE workflow. I. Nagle and G. H. Koenderink gratefully acknowledge the Flagship Healthy Joints, which is (partly) financed by Convergence Health and Technology, for funding.
S. Cardona, M. Peirlinck, and B. Fereidoonnezhad acknowledge the Delft University of Technology for funding. M. Peirlinck also acknowledges support through the NWO Veni Talent Award 20058.

\bibliographystyle{unsrt}
\bibliography{references2}

\renewcommand{\thefigure}{S\arabic{figure}}
\renewcommand{\thesubfigure}{\alph{subfigure}}

\newpage

\renewcommand{\thefigure}{S\arabic{figure}}
\renewcommand{\thesubfigure}{\alph{subfigure}}


\vspace{0.5cm} 

\subsection*{Supplementary Information}

\noindent\hrulefill 
\\
\textbf{Supplementary information for}\\

{\Large{ToFiE, a topology-aware fiber extraction workflow for
3D reconstruction of dense and heterogeneous biological
fibrous network structures from images}}\\

Risa Togo\textsuperscript{a,b}, Sara Cardona \textsuperscript{a}, Irène Nagle\textsuperscript{b}, Gijsje H. Koenderink\textsuperscript{b,*}, Behrooz Fereidoonnezhad\textsuperscript{a,*}, Mathias Peirlinck\textsuperscript{a,*} 
\\

\textsuperscript{a} Department of BioMechanical Engineering, Faculty of Mechanical Engineering, Delft University of Technology, Delft, The Netherlands \\
\textsuperscript{b} Department of Bionanoscience, Kavli Institute of Nanoscience, Delft University of Technology, Delft, The Netherlands \\
\noindent\hrulefill

\subsubsection*{\textbf{NHS ester collagen labeling protocol}}
\label{protocol_ester_labeling}
The methods of Remy et al. and Doyle et al. are adapted in this protocol for labeling bovine atelocollagen type I with DyLight™ 550 NHS Ester \cite{Remy2023InvadopodiaMatrices, Doyle2018FluorescentImaging.}. Bovine atelocollagen type I (Biomatrix Fibricol, Advanced BioMatrix) is neutralized to pH 7.4 in one part 10x phosphate-buffered saline (PBS, pH 7.4, Gibco), 0.1M NaOH, and MilliQ water, to a final volume of 2 mL and a collagen concentration of 5 mg/mL. The solution is mixed in a centrifuge tube on ice by pipetting up and down without introducing bubbles, and then transferred to a glass vial. The pH is checked to be close to 7.4 using pH strips. The solution is incubated in a water bath at 37$^{\circ}\text{C}$ for 90 minutes or until gelation is complete and a smooth opaque gel has formed in the vial. Then the gel is rinsed twice with 5 mL of 1x PBS for ten minutes each. A two-molar excess of DyLight 550 NHS ester dye solution (62262, ThermoScientific™, 10 mg/mL in anhydrous DMSO) is pre-mixed in a buffer consisting of 3 mL of MilliQ and 1 mL of 1M carbonate buffer (1M NaHCO$_3$ adjusted to pH 8.3 with 1M Na$_2$CO$_3$) and added to the gel at room temperature. The volume of dye solution, $v_{\mathrm{Dye}}$ (16 $\mu$L), is calculated based on the mass of collagen in the gel, $m_{\mathrm{Collagen}}$, molecular weight of the dye, $MW_{\mathrm{Dye}}$, 2-molar excess of dye, and the concentration of the dye stock solution, $c_{\mathrm{Dye}}$ (Equation \ref{eq:dye_volume}, \ref{eq:dye_volume_example}) \cite{Doyle2018FluorescentImaging.}. The MW of collagen I, $MW_{\mathrm{Collagen}}$, is theoretically in the range of 130-140 kDa, but variations may exist due to biological reasons such as species, post translational modifications, and experimental factors.

\begin{equation}
\label{eq:dye_volume}
v_{\mathrm{Dye}} =
\frac{m_{\mathrm{Collagen}}}{MW_{\mathrm{Collagen}}}
\times (\text{Dye molar excess})
\times \frac{MW_{\mathrm{Dye}}}{c_{\mathrm{Dye}}}
\end{equation}

\begin{equation}
    \label{eq:dye_volume_example}
    \text{16} = \frac{\text{10}}{\text{130000}} \times \text{2} \times \text{1040} \times \text{100} 
\end{equation}
\\
The dye solution is removed from the gel after one hour and replaced with 10 mL of 50 mM Tris buffer (pH 7.5) to quench the dye reaction for ten minutes. Unbound dye molecules are removed through six sequential washing steps with 5 mL of 1x PBS at room temperature, each time for 30 minutes. In the last wash, all the PBS is discarded. Then 1 mL of cold 20 mM HCl is added followed by the addition of 1M HCl in increments of 10 $\mu$L to bring the pH down to 2, as measured using pH strips, and dissolve the gel. To help the dissolution, the gel is gently agitated for 1-2 hours at 4°C. The final labeled collagen solution is stored at 4°C in the dark. When preparing samples containing a mix of unlabeled and fluorescent labeled collagen, the additional salt present in the labeled collagen solution is accounted for by reducing the volume of 10x PBS added. 

\setcounter{figure}{0} 
\renewcommand{\figurename}{\textbf{Figure}}
\renewcommand{\thefigure}{\textbf{S\arabic{figure}}}

\setcounter{table}{0} 
\renewcommand{\tablename}{\textbf{Table}}
\renewcommand{\thetable}{\textbf{S\arabic{table}}}

\subsubsection*{\textbf{ToFiE user guide}}
\label{User_manual}

\begin{figure}[H]
\centering
\includegraphics[width = \textwidth]{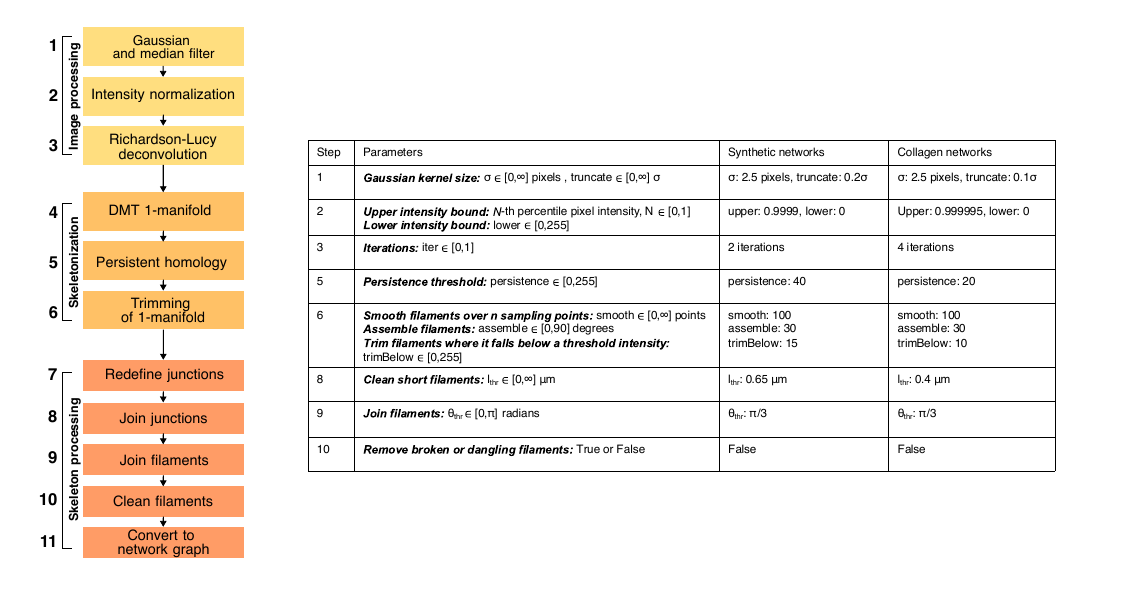}
\caption{\textbf{Overview of the ToFiE workflow parameters and selected values used for network reconstruction.} The schematic and table summarize the user-defined parameters at each step of the workflow, along with the specific values used to reconstruct synthetic networks (Figure \ref{figure2}) and collagen networks (Figure \ref{figure3}). For more details on the parameters in steps 5 and 6, we refer the readers to the DisPerSe manual by Sousbie \cite{Sousbie2011TheImplementation}.}
\label{supp1}
\end{figure}

\begin{figure}[H]
\centering
\includegraphics[width = \textwidth]{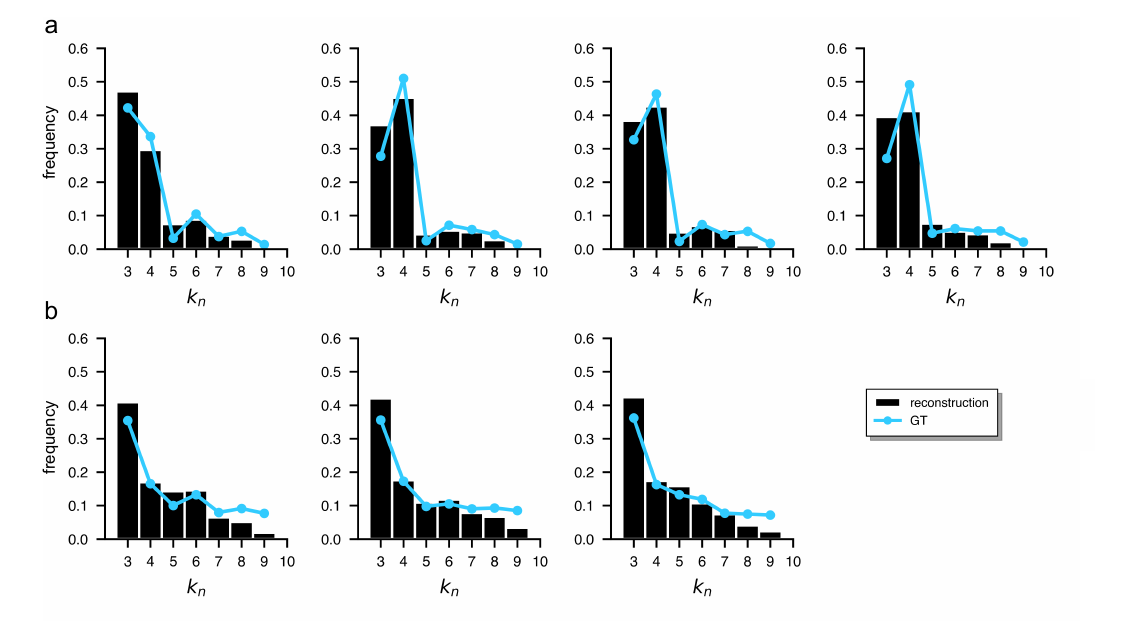}
\caption{\textbf{Dependence of the recall score on the tolerance radius r for a representative synthetic network.} The recall score for nodes with connectivity $k_n=3$ (triangle) or $k_n=4$ (square). Recall is defined as recall $= \frac{TP}{TP+FN}$, quantifying the overlap of nodes with the same connectivity between the reconstruction obtained at the highest SNR condition tested ($\sim 1.9$) and ground truth (GT). Nodes that overlap and have the same $k_n$ are counted as true positives (TP), and GT nodes with no overlapping reconstructed node of the same $k_n$ are false negatives (FN).}
\label{supp2}
\end{figure}

The source code for the ToFiE workflow is publicly available at \url{https://github.com/peirlincklab/ToFiE.git.}. The repository provides detailed installation instructions, example scripts, and a small test dataset with corresponding output files. \\
\\
ToFiE relies on a set of user-defined parameters that allow the workflow to be adapted to different fibrous networks and imaging modalities. Supplementary Figure \ref{supp1} summarizes the parameters associated with each step of the workflow and reports the specific parameter values used for the reconstruction of the synthetic networks shown in Figure \ref{figure2} and the collagen networks shown in Figure \ref{figure3}.

\subsubsection*{}

\begin{figure}[H]
\centering
\includegraphics[width = \textwidth]{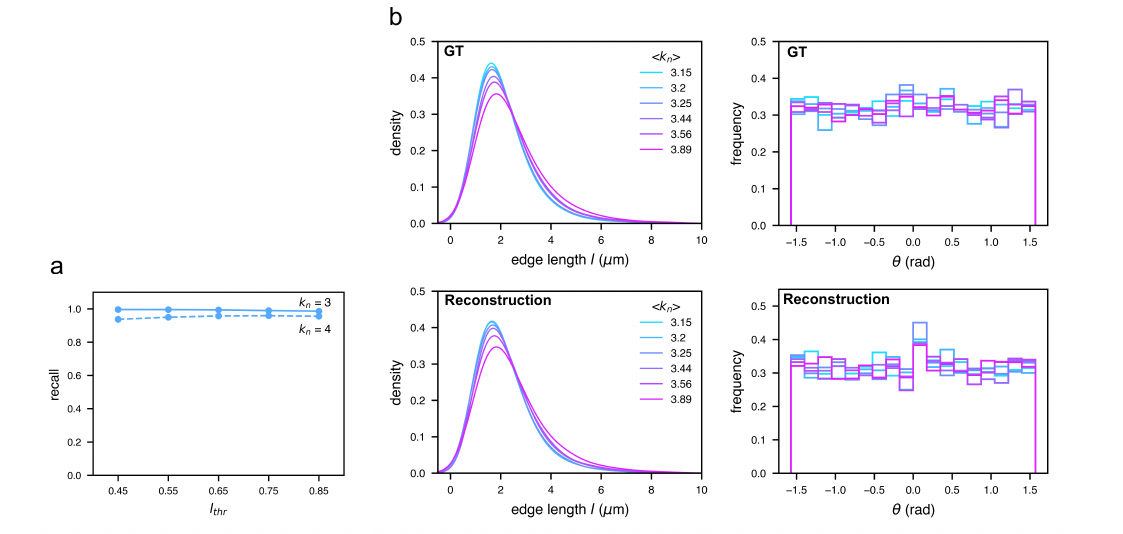}
\caption{\textbf{Sensitivity of network reconstructions to the length threshold $l_{thr}$ and corresponding edge length distributions and azimuthal angle distributions across synthetic networks with average connectivity between three and four.} (a) Recall score as a function of the length threshold for nodes with connectivity $k_n=3$ and $k_n=4$, evaluated using reconstructions obtained at the highest signal-to-noise (SNR) condition ($\sim 1.8$). Error bars represent the standard deviation across N=15 independent networks. (b) Pooled kernel density estimate of edge length distributions and azimuthal angle distributions for the ground truth networks (top row) and the corresponding reconstructions obtained at the highest SNR condition (bottom row), shown for N=2 or N=3 independent networks within each $<k_n>$ group.}
\label{supp3}
\end{figure}

\begin{figure}[H]
\centering
\includegraphics[width = \textwidth]{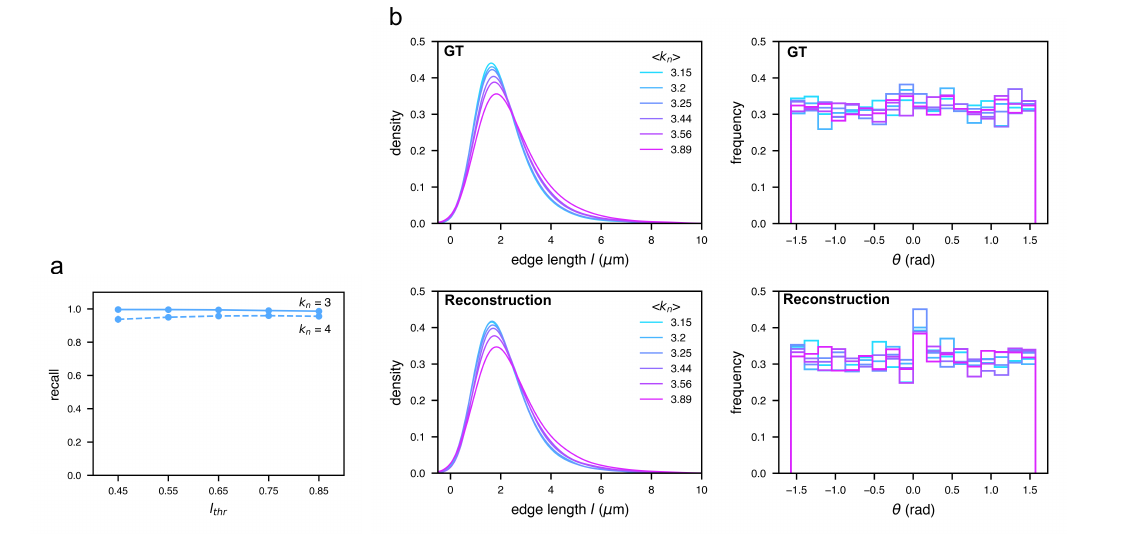}
\caption{\textbf{Edge length and azimuthal angle distributions of example reconstructions across varying image signal-to-noise (SNR).} (Left) Kernel density estimates of edge lengths. (Right) Azimuthal angle distributions. Each row represents a distinct synthetic network and GT stands for ground truth.}
\label{supp4}
\end{figure}

\begin{figure}[H]
\centering
\includegraphics[width = \textwidth]{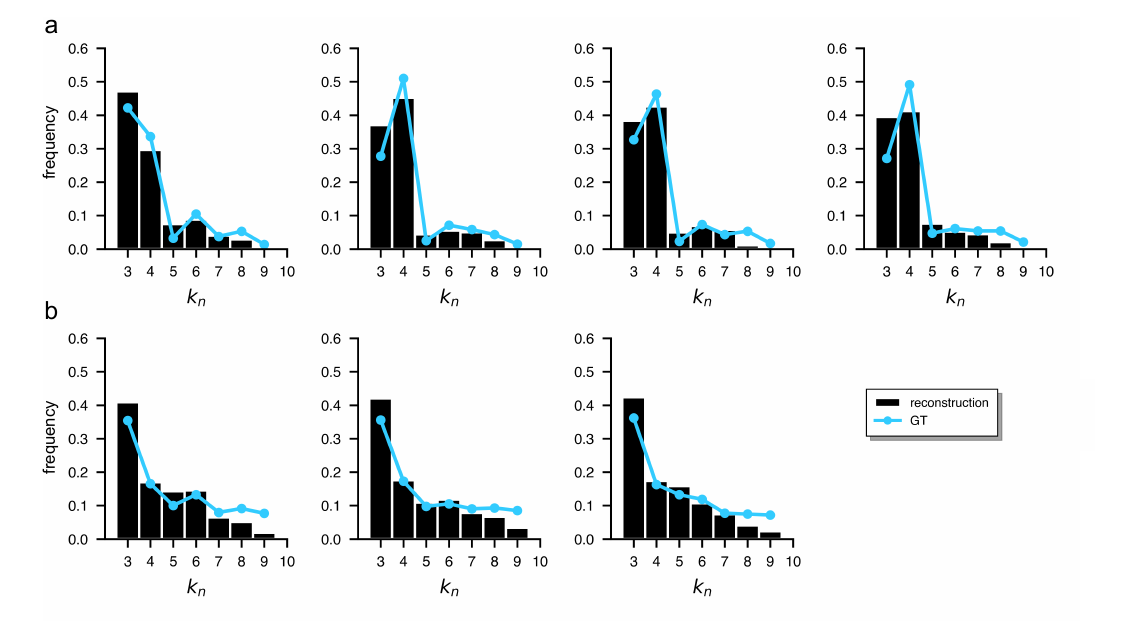}
\caption{\textbf{Comparison of node connectivity distributions between the ground truth (GT) and reconstructed synthetic networks} (at the highest SNR $\sim 1.3-1.5$, $l_{thr}=$0.85 µm). (a) Synthetic networks with average connectivity $<k_n>\sim 4$. (b) Synthetic networks with $<k_n>\sim5$. Each plot represents a distinct synthetic network.}
\label{supp5}
\end{figure}

\begin{figure}[H]
\centering
\includegraphics[width = \textwidth]{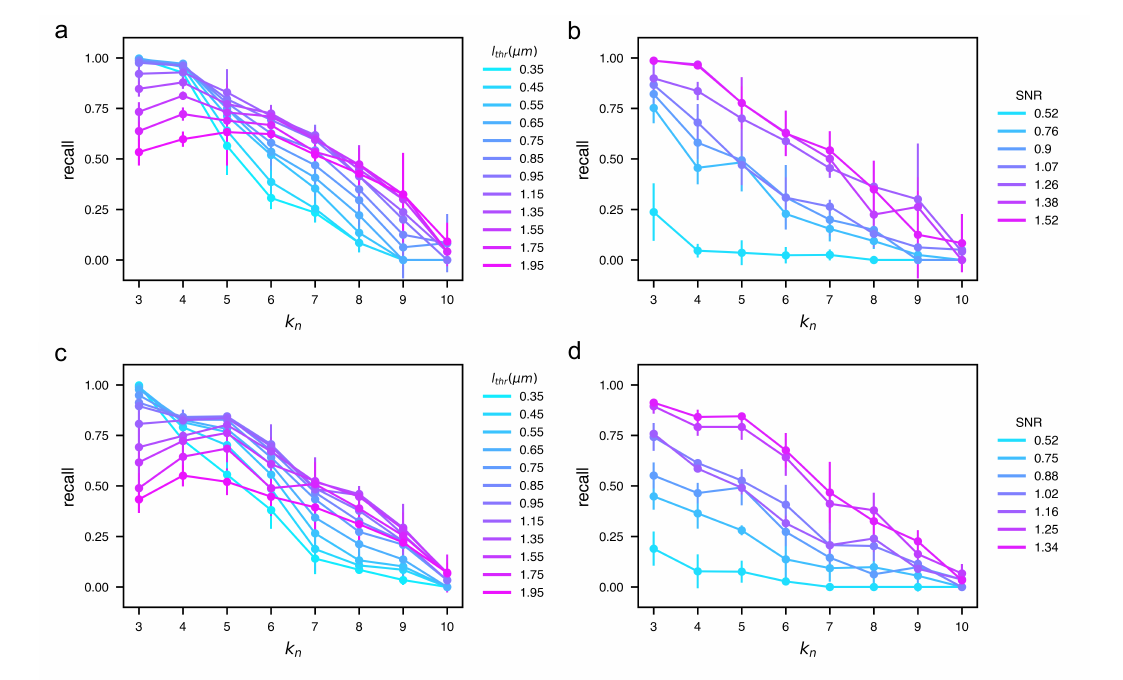}
\caption{\textbf{Sensitivity of network reconstructions to the length threshold $l_{thr}$, and image signal-to-noise ratio (SNR) for synthetic networks with average connectivity $<k_n>$ between four and five.} Top row: Recall score as a function of node connectivity $k_n$ for synthetic networks with $<k_n> \sim 4$, (a) across different length thresholds $l_{thr}$, evaluated at the highest SNR condition ($\sim 1.5$), or (b) across image SNR with $l_{thr}= 0.85$. Error bars represent the standard deviation across $N=4$ independent networks. Bottom row: Recall score as a function of $k_n$ for synthetic networks with $<k_n> \sim 5$, (c) across different length thresholds $l_{thr}$, evaluated at the highest SNR condition ($\sim 1.3$), or (d) across image SNR with $l_{thr}$ $= 0.85$. Error bars represent the standard deviation across $N=3$ independent networks. 
}
\label{supp6}
\end{figure}

\begin{figure}[H]
\centering
\includegraphics[width = \textwidth]{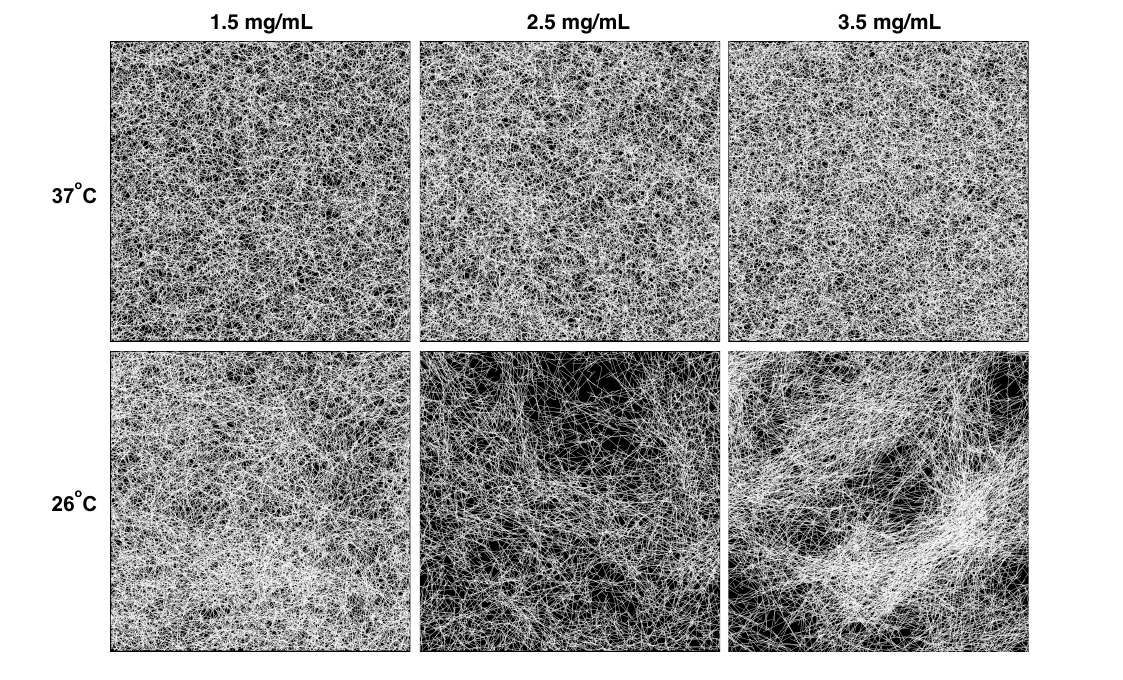}
\caption{\textbf{ToFiE reconstructions of reconstituted collagen I networks imaged by confocal fluorescence microscopy.} Maximum intensity z-projections of the ToFiE reconstructions are shown for collagen gels prepared at different collagen concentrations (1.5, 2.5, 3.5 mg/mL) and polymerization temperatures (37$^{\circ}\text{C}$, 26$^{\circ}\text{C}$). Each frame represents a lateral field of view of 111.94 $\mu$m over a depth of 4.95 $\mu$m.}
\label{supp7}
\end{figure}

\begin{table}[H]
\centering
\begin{tabular}{|c|c|c|c|c c c c c|}
\hline
\textbf{Condition} & \textbf{$\overline{l}(\mu m)$} & \textbf{$v$} & $R^2$ &   &  & \textbf{p-value} &  & \\
\hline 
37$^{\circ}\text{C}$, 1.5 mg/mL (N=1) & 2.13 & 1.07 & 0.993 & ref. & & ref. & & \\
37$^{\circ}\text{C}$, 2.5 mg/mL (N=2) & 1.98 & 0.92 & 0.995 & * & & & ref. &\\
37$^{\circ}\text{C}$, 3.5 mg/mL (N=2) & 1.83 & 0.78 & 0.995& *** & & & & ref. \\
26$^{\circ}\text{C}$, 1.5 mg/mL (N=2) & 2.01 & 0.92 & 0.998 & & ref. 
 & $^{\dag\dag\dag}$ & &\\
26$^{\circ}\text{C}$, 2.5 mg/mL (N=2) & 2.39 & 1.09 & 0.996 & & *** & & $^{\dag\dag\dag}$ &\\
26$^{\circ}\text{C}$, 3.5 mg/mL (N=2) & 2.12 & 1.06 & 0.998 & & n.s. & & & $^{\dag\dag\dag}$\\
\hline
\end{tabular}
\caption{\textbf{Edge length distribution of collagen network reconstructions across different polymerization temperatures and collagen concentrations.} Edge lengths $l$ are binned into 30 equally spaced bins in the range $ [0, 20]$ $\mu$m and fitted to a log-normal distribution $f_{log-normal}(l, \overline{l},v) = \frac{1}{\overline{l}\sqrt{2\pi\xi^{2}}} e^{-\frac{(\lambda -\ln{l})^{2}}{2\xi^{2}}}$, where $\xi^{2} = \ln{v+1}$, and $\lambda = \ln{\overline{l}} - \frac{\xi^{2}}{2}$ with parameters $\overline{l}$ and $v$ as the mean and normalized variance of the distribution \cite{Lindstrom2010BiopolymerProperties}.
Significant differences between distributions are tested using the Mann-Whitney U-test. Star symbols (*) indicate a significant difference from the reference of 1.5 mg/mL concentration within the same temperature condition (*: p $\le $0.05, ***: p $\le $0.0005, n.s.: not significant). Dagger symbols ($\dag$) indicate significant difference between the temperatures 37$^{\circ}\text{C}$ and 26$^{\circ}\text{C}$ at the same collagen concentration ($\dag\dag\dag$: p $\le$0.0005). }
\label{stable1}
\end{table}

\begin{figure}[H]
\centering
\includegraphics[width = \textwidth]{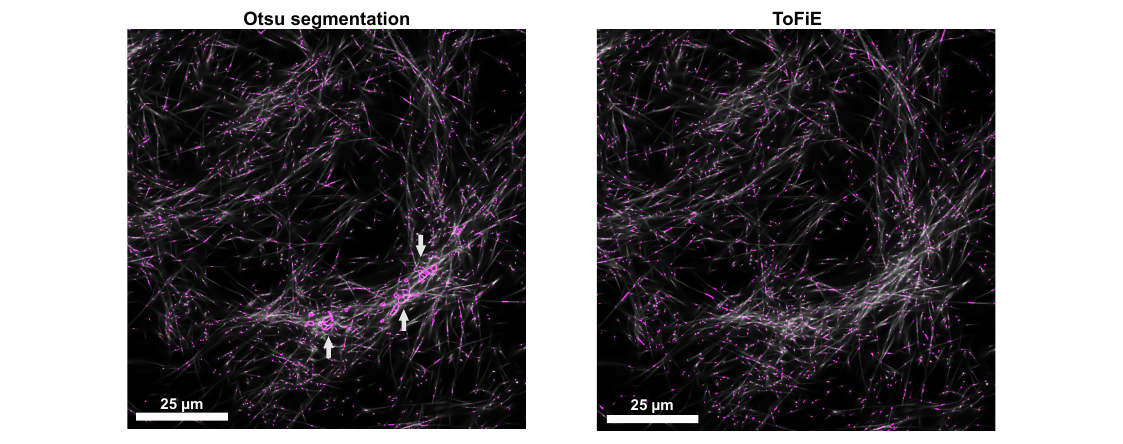}
\caption{\textbf{ToFiE reconstruction and Otsu segmentation superimposed on confocal fluorescence image of a dense and heterogeneous collagen network.} The image (gray) depicts a 3.5 mg/mL collagen network polymerized at 26$^{\circ}\text{C}$ across a 0.1 $\mu$m depth along the z-axis. For Otsu segmentation (magenta, left), Otsu thresholding is applied to the preprocessed 3D confocal image (after the first step of ToFiE) in Fiji, with the lower threshold value of 50 and upper threshold value of 255. The thresholded binary mask is thinned to a skeleton by applying Skeletonize3D in Fiji \cite{Lee1994BuildingAlgorithms}. The ToFiE reconstruction (magenta, right) visually traces out the fibers in the image. The ToFiE reconstruction and Otsu segmentation are both dilated once using morphological dilation for easier visualization. White arrows point at artifacts in the Otsu skeleton, arising from ill-defined topology in the binary mask, especially in regions with densely bundled fibers.
}
\label{supp8}
\end{figure}

\begin{figure}[H]
\centering
\includegraphics[width = \textwidth]{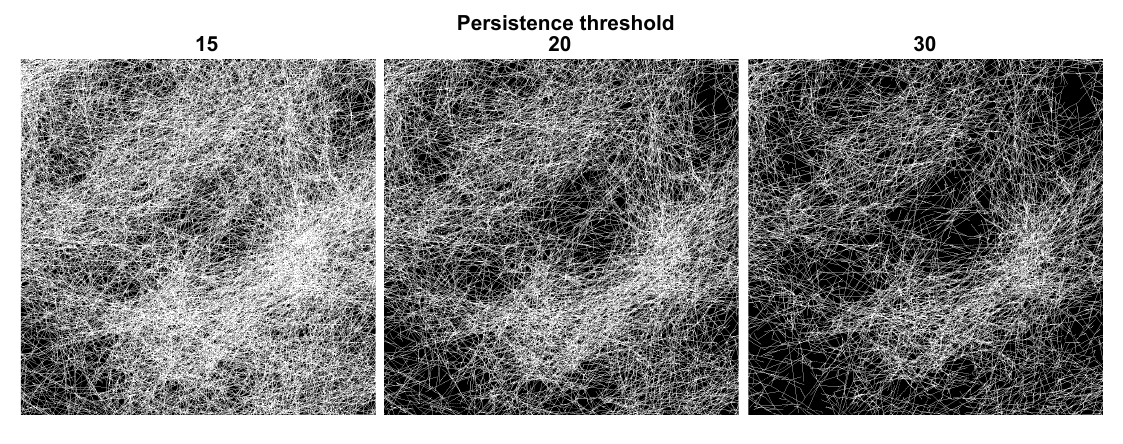}
\caption{\textbf{ToFiE reconstruction for varying persistence thresholds demonstrates topology-preserving nature.} Maximum intensity z-projections of the ToFiE reconstruction of a 3.5 mg/mL collagen network polymerized at 26$^{\circ}\text{C}$, are shown for varying persistence thresholds (15, 20, 30). As the persistence threshold increases, less persistent topological features (quantified as the intensity difference of critical 0- and 1-simplex pairs) are progressively removed. The remaining critical pairs are merged together for a simplified connected topological structure. The persistence threshold of 20 is chosen for the reconstruction of confocal images of collagen networks.}
\label{supp9}
\end{figure}

\section*{Supplementary videos}
\setcounter{figure}{0}
\renewcommand{\figurename}{\textbf{Supplementary Video}}
\renewcommand{\thefigure}{\textbf{S\arabic{figure}}}

\begin{figure}[H]
\centering
\caption{\textbf{Overlay video showing the Otsu segmentation with a low Otsu threshold superimposed on confocal fluorescence z-slices of a dense and heterogeneous collagen network.} The images (gray) depict a 3.5 mg/mL collagen network polymerized at 26$^{\circ}\text{C}$, with each frame representing 0.1 $\mu$m depth along the z-axis. The images are the same one shown in all three Supplementary Videos. For Otsu segmentation (in magenta), Otsu thresholding is applied in Fiji to the preprocessed confocal image (after the first step of ToFiE) with the lower threshold value of 50 and upper threshold value of 255. The thresholded binary mask is thinned to a skeleton by applying Skeletonize3D in Fiji \cite{Lee1994BuildingAlgorithms}. The Otsu skeleton is dilated once using morphological dilation for easier visualization in the video.
}
\label{suppvideo1}
\end{figure}

\begin{figure}[H]
\centering
\caption{\textbf{Overlay video showing the Otsu segmentation with a high Otsu threshold superimposed on confocal fluorescence z-slices of a dense and heterogeneous collagen network.} The images (gray) depict a 3.5 mg/mL collagen network polymerized at 26$^{\circ}\text{C}$, with each frame representing 0.1 $\mu$m depth along the z-axis. The images are the same one shown in all three Supplementary Videos. For Otsu segmentation (in magenta), Otsu thresholding is applied in Fiji to the preprocessed confocal image (after the first step of ToFiE) with the lower threshold value of 130 and upper threshold value of 255. The thresholded binary mask is thinned to a skeleton by applying Skeletonize3D in Fiji \cite{Lee1994BuildingAlgorithms}. The Otsu skeleton is dilated once using morphological dilation for easier visualization in the video.
}
\label{suppvideo2}
\end{figure}

\begin{figure}[H]
\centering
\caption{\textbf{Overlay video showing the ToFiE reconstruction superimposed on confocal fluorescence z-slices of a dense and heterogeneous collagen network.} The images (gray) depict a 3.5 mg/mL collagen network polymerized at 26$^{\circ}\text{C}$, with each frame representing 0.1 $\mu$m depth along the z-axis. The images are the same one shown in all three Supplementary Videos. The ToFiE workflow reconstructs the network from the images (magenta), and is dilated once using morphological dilation for easier visualization in the video.}
\label{suppvideo3}
\end{figure}


\end{document}